\documentclass[10pt,twocolumn]{article}

\usepackage[
  top=2cm,
  bottom=2cm,
  left=1.8cm,
  right=1.8cm,
  columnsep=0.6cm
]{geometry}

\usepackage{amsmath, amssymb, amsthm, mathrsfs,bm}
\usepackage{url}
\usepackage{orcidlink}
\usepackage{hyperref}
\usepackage{placeins}  
\usepackage{graphicx} 
\usepackage[sorting=none]{biblatex}
\usepackage{titling}
\usepackage{fancyhdr}
\usepackage{amsmath} 
\usepackage{booktabs} 
\usepackage{amsfonts} 
\usepackage{mathtools} 
\usepackage{float} 
\usepackage{tabularx}
\usepackage{array}

\usepackage{xcolor}
\usepackage{subcaption}

\title{Data-Driven Generation of Neutron Star Equations of State Using Variational Autoencoders}

\author{
\large{Alex Ross\,\:\orcidlink{0009-0005-8103-5823}\textsuperscript{1}},
\large{Tianqi Zhao\,\:\orcidlink{0000-0003-4704-0109}\textsuperscript{1},
Sanjay Reddy\,\:\orcidlink{0000-0003-3678-6933}\textsuperscript{1}}\\
{\small \textsuperscript{1}
Institute for Nuclear Theory, University of Washington, Seattle, WA, 98105, USA}
}
\date{\today}

\begin{document}

\twocolumn[
\begin{@twocolumnfalse}
\maketitle
\begin{abstract}
We develop a machine learning model based on a structured variational autoencoder (VAE) framework to reconstruct and generate neutron star (NS) equations of state (EOS). The VAE consists of an encoder network that maps high-dimensional EOS data into a lower-dimensional latent space and a decoder network that reconstructs the full EOS from the latent representation. The latent space includes supervised NS observables derived from the training EOS data, as well as latent random variables corresponding to additional unspecified EOS features learned automatically. Sampling the latent space enables the generation of new, causal, and stable EOS models that satisfy astronomical constraints on the supervised NS observables, while allowing Bayesian inference of the EOS incorporating additional multimessenger data, including gravitational waves from LIGO/Virgo and mass and radius measurements of pulsars. Based on a VAE trained on a Skyrme EOS dataset, we find that a latent space with two supervised NS observables, the maximum mass \(M_{\max}\) and the canonical radius \(R_{1.4}\), together with one latent random variable controlling the EOS near the crust--core transition, can already reconstruct Skyrme EOSs with high fidelity, achieving mean absolute percentage errors of approximately \(0.15\%\) for \(M_{\max}\) and \(R_{1.4}\) derived from the decoder-reconstructed EOS.
\end{abstract}
\vspace{0.5cm}
\end{@twocolumnfalse}
]

\newpage

\section{Introduction}
The cold neutron star (NS) equation of state (EOS) describes the relationship between the pressure and the energy density at zero temperature in stable matter within a NS \cite{SumiyoshiEOS}. The EOS is well constrained at low densities near the crust by nuclear theory and experiment, and at asymptotically high densities beyond those realized in NSs by perturbative QCD. However, the intermediate-to-high density EOS relevant to NS cores remains highly uncertain \cite{2017arXiv171008220D} \cite{2023ApJ...950...77H} \cite{2018RPPh...81e6902B}. Matter at extreme densities (up to twice nuclear saturation density) can be described by \textit{ab initio}
nuclear theory frameworks
such as Chiral Effective Field Theory ($\chi$EFT) \cite{2021ARNPS..71..403D}. Additionally, phenomenological models such as relativistic mean-field and the Skyrme nuclear models can be used to extrapolate nuclear interactions to higher densities, but these mean field approaches lack higher order quantum corrections to the many-body forces, and the associated theoretical uncertainties are not quantifiable. In this regime, phase transitions may play an important role due to the possible emergence of exotic degrees of freedom (e.g., hyperons or deconfined quarks) \cite{RMF1} \cite{RMF2}. As a result, different theoretical frameworks can produce EOS that are consistent with low-density constraints yet diverge significantly at higher densities, leading to large variations in predicted NS masses, radii, and tidal deformabilities. This uncertainty motivates the development of flexible, data-driven approaches that can systematically explore the space of physically admissible EOS while remaining consistent with both theoretical constraints and astrophysical observations. Such data driven approaches have, as of late, largely included machine learning (ML) as a way of rapidly learning, condensing, and scanning a large parameter space. 

In particular, many works focus on the use of the Gaussian Process (GP) conditioned on theoretical nuclear models \cite{GP} \cite{GP2} \cite{GP3}. These studies focus on conditioning the GP on observations to get posterior probability distributions for EOS parameters (like radius, maximum mass). Works such as \cite{2023ApJ...950...77H} and \cite{EoSML} used ML to study the relationship between NS observables and the underlying parameters. In \cite{2023ApJ...950...77H}, the authors employ the variational autoencoder (VAE) deep learning framework to generate physically valid NS EOS, incorporating observational data to constrain the possible parameter space. In this work, we expand on this VAE-based approach by systematically analyzing the physical structure and interpretability of the learned latent space, quantifying how individual latent directions map to variations in the decoded EOS and NS observables. We train a fully generative VAE model using data derived from the Skyrme nuclear model, using this input data to calculate theoretical maximum stable NS mass configuration, and the radius of a 1.4 M$_\odot$ NS which will be used as supervised latent observables during model training. Once we have a fully trained and generative model, we vary parameters to decode the behavior of the NS EOS when controlled by 3 parameters rather than the 10 parameters required by the Skyrme model. By varying these parameters, we can control the behavior of the generated EOS. For example, for a given EOS, scaling the value of the parameter corresponding to the maximum stable mass configuration of a NS should provide a stiffer EOS at high densities. Scaling the parameter corresponding to the radius of a 1.4 $M_\odot$ NS $(R_{1.4})$ should provide identical EOS for each multiple of $R_{1.4}$ except around twice nuclear saturation density, where the EOS is known to be more sensitive to changes in $R_{1.4}$ \cite{r14sensitivity} and each EOS will differ. We additionally can vary the learned latent parameter(s) $z_i$ to test and determine which density region of the EOS is being controlled by said parameter.

\section{Methods}
\subsection{Energy Density Functional}\label{subsec:skyrme}
Before we discuss the VAE, it is necessary to formalize the data that the framework will be trained on. The VAE is trained on a diverse candidate EOS dataset derived from the Skyrme nuclear model. The Skyrme Hamiltonian density for infinite nuclear matter is written as follows \cite{Tianqi}:
\begin{equation}
\varepsilon_{N} =
\frac{k_{F n}^{5}}{10\pi^{2} m_n^{*}}
+
\frac{k_{F p}^{5}}{10\pi^{2} m_p^{*}}
+
H_{\mathrm{pot}}(n_n, n_p),
\label{eq:Hsk}
\end{equation}
where the first two terms are the kinetic contributions for neutrons and protons including the
effective mass, and $H_{\mathrm{pot}}$ is the potential term given by
\begin{equation}
\begin{aligned}
H_{\mathrm{pot}} &=
\frac{1}{2} n^{2} t_{0}
\left( 1 + \frac{x_{0}}{2} \right)
-
\frac{1}{2} (n_n^{2} + n_p^{2}) t_{0}
\left( \frac{1}{2} + x_{0} \right)
\\
&\quad
+
\frac{1}{24} n^{\gamma} t_{3}
\left[
n^{2} (2 + x_{3})
-
(n_n^{2} + n_p^{2})(1 + 2x_{3})
\right] .
\end{aligned}
\label{eq:Hpot}
\end{equation}
The neutron and proton effective masses are density dependent and given by
\begin{equation}
\begin{aligned}
\frac{m_{n/p}}{m_{n/p}^{*}} &= 1
+ \frac{m_{n/p}}{4}
\Big\{
n \left[ t_{1}(2 + x_{1}) + t_{2}(2 + x_{2}) \right]
\\
&\qquad
+ n_{n/p}
\left[ -t_{1}(1 + 2x_{1}) + t_{2}(1 + 2x_{2}) \right]
\Big\} .
\end{aligned}
\label{eq:meff}
\end{equation}

The total energy per baryon is
\begin{equation}
E(n_B, x) = \varepsilon_N(n_B, x)/n_B + E_e(n_B, x),
\end{equation}
where proton ratio $x=\frac{n_p}{n_B}$,  the energy density of nuclear matter $\varepsilon_N$ is given in Eq. (\ref{eq:Hsk}), and $E_e$ is the energy 
of a relativistic electron gas with $n_e = xn_B$ (charge neutrality):
\begin{equation}
\begin{aligned}
E_e(n_e)
&=
\frac{1}{\pi^2 n_e}
\int_0^{k_{Fe}}
\sqrt{k^2 + m_e^2}\, k^2 \, dk
\approx \frac{3}{5} k_{Fe}, \\
k_{Fe}
&=
(3\pi^2 n_e)^{1/3}.
\end{aligned}
\end{equation}

A cold $\beta$-equilibrated EOS is obtained by minimizing the energy per baryon with respect to the proton fraction at fixed baryon density,
\begin{equation}
    \frac{\partial E(n_B,x)}{\partial x}|_{n_B,x=x_{eq}(n_B)}=0.
\end{equation}
This defines the equilibrium proton fraction $x_{\rm eq}(n_B)$. The corresponding barotropic EOS is specified by
\begin{equation}
\begin{aligned}
\varepsilon(n_B)
&=
n_B\,E\!\left(n_B,x_{\rm eq}(n_B)\right), \\
p(n_B)
&=
n_B^2\,\frac{d}{dn_B}
E\!\left(n_B,x_{\rm eq}(n_B)\right).
\end{aligned}
\end{equation}

which determine the structure of NS as detailed in Sec. \ref{subsec:TOV}.

Although such an EOS formally extends to vanishing density and pressure, it should not be applied beyond the uniform core of a neutron star. Uniform nuclear matter becomes unstable to clustering below the crust-core transition density $n_{cc}$. $n_{cc}$ can be estimated using the thermodynamic spinodal condition, which requires the Hessian matrix of the energy per baryon with respect to $(n_B,x)$ to be positive definite,
\begin{equation}
\begin{aligned}
\det[H] &=
\frac{\partial^{2} E(n_B,x)}{\partial n_B^{2}}
\frac{\partial^{2} E(n_B,x)}{\partial x^{2}}
-
\left(
\frac{\partial^{2} E(n_B,x)}
     {\partial n_B\,\partial x}
\right)^{2}
> 0 .
\end{aligned}
\label{eq:stability}
\end{equation}
For matter in beta equilibrium, the stationarity condition
$\partial E / \partial x = 0$ holds and the curvature
$\partial^{2} E / \partial x^{2} > 0$ ensures stability with respect to composition fluctuations. In this case, the spinodal condition reduces to
\begin{equation}
\frac{\partial^{2} E(n_B,x_{\rm eq})}{\partial n_B^{2}} > 0 .
\label{eq:nBcc_condition}
\end{equation}
At density below $n_{cc}$, the above stability condition Eq. (\ref{eq:nBcc_condition}) is violated, which marks the onset of an instability of uniform nuclear matter toward clustering into nuclei. To calculate $n_{max}$, or the density of the maximum stable NS mass configuration, we first solve the TOV equations as given in Sec. \ref{subsec:TOV} to acquire the maximum stable mass point. We then evaluate the $n_{max}$ at this point and use it as part of our boundary condition dataset. We also take the pressure $P_{max}$ and energy density $\varepsilon_{max}$ at this point to fully constrain the upper boundary on our EOS. To compute the pressure and energy density at the core-crust transition density ($P_{cc}$ and $\varepsilon_{cc}$, respectively), we simply evaluate the EOS at the density $n_{cc}$. All boundary values discussed in this work are summarized in Tab. ~\ref{tab:input_data}.

\subsection{Training Data}\label{subsec:vae-input}
We use the Skyrme model discussed in Sec. \ref{subsec:skyrme} to derive our input EOS training data. The structure of the input data is an array of dimensions $\mathbf{[m,\,107]}$, where $m$ is the number of distinct EOS. The first 101 columns are comprised of sound speed data, with the sound speed $c_s^2$ given by the following equation:
\begin{equation}
  c_s^2(\zeta)=\frac{dp}{d\varepsilon}, \qquad \zeta \equiv \ln p,
  \label{eq:csdef}
\end{equation}
 The final 6 columns that comprise the input data are boundary conditions that set critical components of the NS structure. This includes theoretical maximum limits of baryon number, energy density, and pressure at the core of a stable NS, as well as the baryon number, energy density, and pressure at the core-crust transition. These core-crust transition values are used to smoothly join the target core EOS to the crust EOS predicted by the Skyrme nuclear model. The boundary values are calculated as summarized in Sec. \ref{subsec:skyrme}. The entire input data array is then log-scaled to stabilize the training process. To acquire our pressure values, we define a logarithmically spaced grid as follows;
\begin{equation}
    p_i = \exp\!\left[\ln p_{cc} 
        + \frac{i}{N}\big(\ln p_{\mathrm{max}} - \ln p_{cc}\big)\right],
    \label{eq:pressuregrid}
\end{equation}

with $N=101$ points spanning the interval between the core-crust 
transition pressure $p_{cc}$ and the pressure of a stable maximum-mass configuration NS $p_{\mathrm{max}}$. The Skyrme nuclear model is used to calculate $p_{cc}$ and $p_{\mathrm{max}}$. For each pressure value $p_i$, the corresponding energy density $\varepsilon(p_i)$ is obtained from the SLy4 EOS table (a specific parametrization of the Skyrme effective nucleon-nucleon interaction \cite{sly4} \cite{sly4-2}). We then numerically compute the sound speed via a finite-difference scheme using Eq. (\ref{eq:csdef}). This data is then split into smaller datasets that are used for different purposes during training. The training split is summarized in Tab. \ref{tab:data_splits}. The bulk of the data is allocated to the training dataset, which is used to minimize the loss function during the VAE training process. The validation dataset serves to tune hyperparameters and decide when training should be terminated without biasing the final result. The test dataset is not used at all during training to calculate gradients or update weightings, and is necessary to evaluate final model performance and report unbiased reconstruction accuracy.

\begin{table}[t]
\centering
\caption{VAE Boundary Value Data}
\label{tab:input_data}
\small
\begin{tabularx}{\columnwidth}{l >{\centering\arraybackslash}X >{\centering\arraybackslash}X}
\toprule
 & \textbf{Core--Crust Transition} & \textbf{Maximum} \\
\midrule
Baryon number  & $n_{cc}$          & $n_{max}$ \\
Energy density & $\varepsilon_{cc}$& $\varepsilon_{max}$ \\
Pressure       & $P_{cc}$          & $P_{max}$ \\
\bottomrule
\end{tabularx}
\end{table}

\subsection{Reconstructing EOS from Data}
Our primary goal in this work is to create new candidate EOS using sound speed profiles predicted by our VAE. We use the boundary values discussed in Sec. \ref{subsec:skyrme} and \ref{subsec:vae-input} to define the physically valid range of each EOS sample, allowing the VAE to learn only within the causal and thermodynamically stable region between the crust–core transition and the maximum central density. To calculate the NS EOS with the sound speed, we use the following formulation. Starting from Eq. (\ref{eq:csdef}), we note that
\begin{equation}
  p = e^{\zeta}, \qquad dp = e^{\zeta}\, d\zeta.
  \label{eq:pzeta}
\end{equation}
Substituting \eqref{eq:pzeta} into \eqref{eq:csdef} gives
\begin{equation}
  d\varepsilon = \frac{dp}{c_s^2(\zeta)} = \frac{e^{\zeta}}{c_s^2(\zeta)}\, d\zeta.
  \label{eq:deps}
\end{equation}
Integrating from a reference state \((p_0,\varepsilon_0)\) to \((p,\varepsilon)\),
\begin{equation}
  \varepsilon(p) = \varepsilon_0 + \int_{\ln p_0}^{\ln p} \frac{e^{\zeta}}{c_s^2(\zeta)}\, d\zeta,
  \label{eq:eps_of_p_zeta}
\end{equation}
which is equivalently, in pressure space,
\begin{equation}
  \varepsilon(p) = \varepsilon_0 + \int_{p_0}^{p} \frac{1}{c_s^2(\ln p')}\, dp'
  \label{eq:eps_of_p}
\end{equation}

where $\varepsilon(p)$ is the EOS expressed in terms of the energy density as a function of pressure. In general, Eqs. \eqref{eq:eps_of_p_zeta} and \eqref{eq:eps_of_p} have no closed form unless \(c_s^2\) has a simple analytic expression; otherwise they should be evaluated numerically.

The input and output data consists of a set of sampled sound-speed values 
$\{c_s^2(p_i)\}_{i=1}^{101}$, where $p_i$ is calculated using Eq. \eqref{eq:pressuregrid}, together with thermodynamic boundary 
conditions at the core-crust transition and at the maximum density.  
These sound speed values are spline-interpolated to obtain a continuous function 
$c_s^2(\ln p)$.  We then solve Eq. (\ref{eq:eps_of_p}) using this function to acquire the corresponding $\varepsilon(p)$. The baryon density $n_B(p)$ is obtained 
by integrating the ODE
\begin{equation}
\frac{d\ln n_B}{d\ln p}
   =\frac{1}{\varepsilon+p}\frac{d\varepsilon}{d\ln p}.
\end{equation}
The resulting EOS segment $\{n_B(p),\varepsilon(p), p\}$ is then joined to the SLy4 crust EOS. 

\subsection{The TOV Equations}\label{subsec:TOV}
The TOV equations describe the structure of a static NS in hydrodynamical equilibrium \cite{2025EPJA...61...55C} \cite{Tolman} \cite{Oppy}. The equations require an input EOS to solve, unless reformulated in a dimensionless form, where the pressure and energy density are expressed as polynomial functions of a reduced radial coordinate as in \cite{2025EPJA...61...55C}. However, employing an explicit input EOS remains the standard approach for modeling NS structure. The TOV equations for a spherically symmetric body in static equilibrium are given by, 

\begin{equation}
\begin{aligned}
\frac{dp(r)}{dr}
&= -\,\frac{G\,\varepsilon(p)\,m(r)}{r^2}
 \left(1 + \frac{p(r)}{\varepsilon(p)c^2}\right)
 \left(1 + \frac{4\pi r^3 p(r)}{m(r)c^2}\right) \\
&\quad\times
 \left(1 - \frac{2G m(r)}{r c^2}\right)^{-1}
\end{aligned}
\label{eq:tov_dpdr}
\end{equation}

and, 
\begin{align}
\frac{dm(r)}{dr}
&= 4\pi r^2 \,\frac{\varepsilon(p)}{c^2}, \label{eq:tov_dmdr}
\end{align}
where $\varepsilon(p)$ is the energy density as a function of pressure, and $p(r)$ is the pressure as a function of radius. In this work, we solve the TOV equations at two separate stages, prior to and post-training. Prior to training, we use input EOS as calculated using the Skyrme model to compute theoretically possible NS maximum masses and radii of a 1.4$M_{\odot}$ NS with the TOV equations, which are used as supervised latent variables during training. We then solve the equations again post-training, using the sound speed profile $c_s^2(p)$ generated by the VAE to solve Eq. (\ref{eq:eps_of_p}) and using the corresponding EOS as input to the TOV equations. We pick a central density and calculate the central pressure for a given EOS. Our TOV solver uses the SciPy LSODA method \cite{Petzold1983LSODA} to integrate Eq. (\ref{eq:tov_dpdr}) and Eq. (\ref{eq:tov_dmdr}) outward from the stellar center, where
$m=0$, $r=0$, $\varepsilon=\varepsilon_c$, and $p=p_c$, to the surface
$r=R$, defined by the condition $p(R)=0$, at which point $m(R)=M$. Repeating this process across a range of central pressures $p_c$ produces the full mass-radius (MR) relation, $R(M)$. Once the high-density portion is computed, it is smoothly joined to the SLy4 crust EOS.

\subsection{The Structured Variational Autoencoder}\label{subsec:SVAE}
\begin{figure*}[t]
  \centering
  \includegraphics[width=\linewidth]{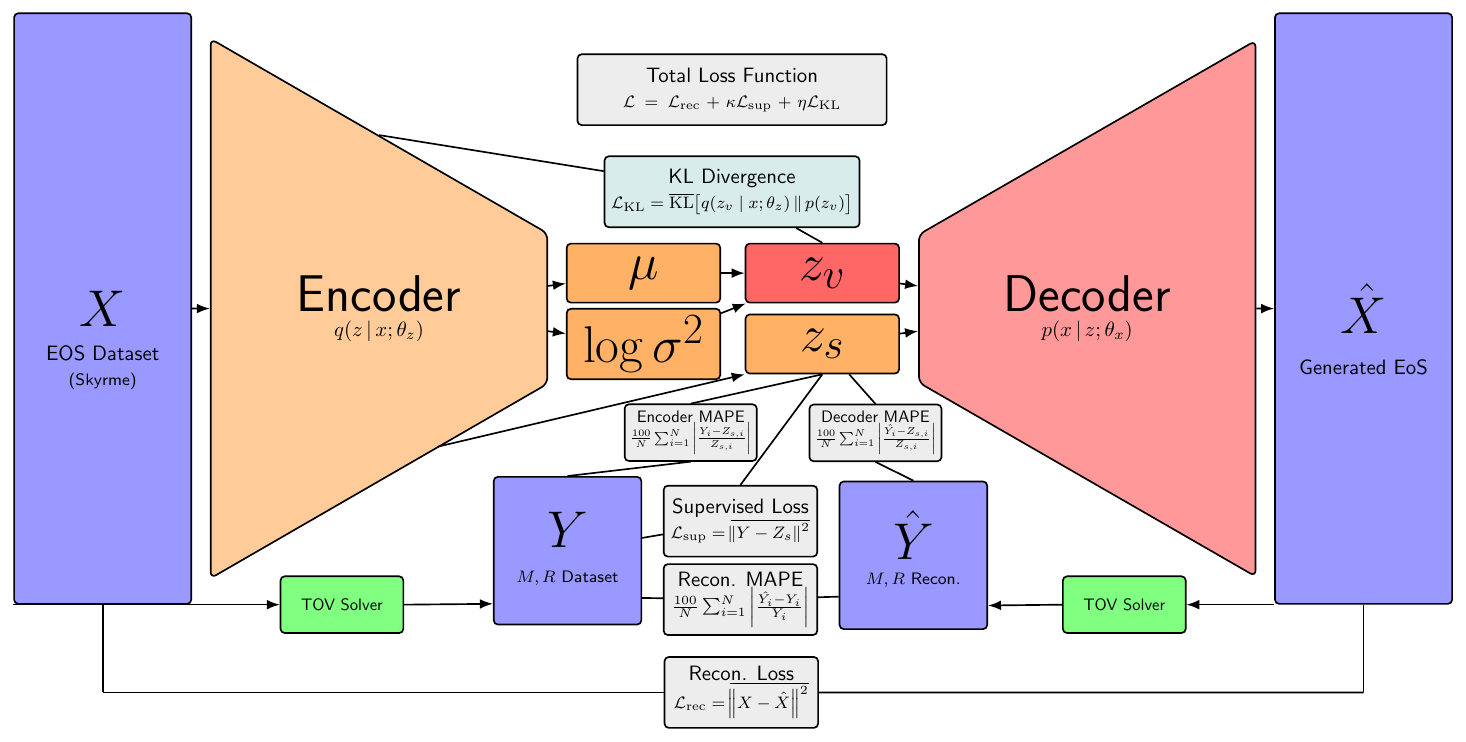}
  \caption{The Variational Autoencoder Framework.}
  \label{fig:VAE}
\end{figure*}
Considering a distribution for EOS $x$ and an underlying latent parameter space $z$, the original intractable problem is fundamentally about understanding all four key components in conditional probability: 
the prior $p(z)$, the likelihood $p(x \mid z)$, the posterior $p(z \mid x)$, 
and the marginal likelihood $p(x)$, satisfying the Bayes’ rule $p(x \mid z)p(z)=p(z \mid x)p(x)$. To make the problem tractable, we introduce a neural network to approximate these probability distributions.  Given an EOS dataset $X=\{x_n\}_{n=1}^N$ and a chosen prior $p(z)$, our modeling goal is to learn the parameters of the likelihood $p(x\mid z)\approx p(x\mid z;\theta_x)$ and thereby characterize the posterior $p(z\mid x)\approx q(z\mid x;\theta_z)$ defined by neural network parameters $\theta_x$ and $\theta_z$. To approach this, we aim to maximize the marginal likelihood of the data $\mathcal{L}$,
\[
\mathcal{L}(\theta_x)=\sum_{n=1}^{N} \log \int p(x_n \mid z;\theta_x)\,p(z)\,dz.
\]
which requires integrating the product of the likelihood and the prior over the latent variables and summing over all data examples. This expression represents the probability (evidence) of the data under the model. Taking the posterior $q(z\mid x;\theta_z)$ as the variational distribution, we can rewrite
\begin{align}
\mathcal{L}(\theta_x,\theta_z)
&= \sum_{n=1}^{N}
\log \int q(z \mid x_n;\theta_z)
\frac{p(x_n \mid z;\theta_x)\,p(z)}
     {q(z \mid x_n;\theta_z)}
\, dz \notag \\
&\ge \mathcal{L}_{\mathrm{ELBO}},
\\[0.5em]
\mathcal{L}_{\mathrm{ELBO}}
&= \sum_{n=1}^{N}
\int q(z \mid x_n;\theta_z)\,
\log\!\left[
\frac{p(x_n \mid z;\theta_x)\,p(z)}
     {q(z \mid x_n;\theta_z)}
\right] dz \notag \\
&= \sum_{n=1}^{N}
\int q(z \mid x_n;\theta_z)\,
\log p(x_n \mid z;\theta_x)\,dz \notag \\
&\quad
- \sum_{n=1}^{N}
\mathrm{KL}\!\left(
q(z \mid x_n;\theta_z)\,\|\,p(z)
\right).
\label{eq:elbo}
\end{align}

where Jensen's inequality is used to derive the evidence lower bound (ELBO) in Eq. (\ref{eq:elbo}). This objective is jointly optimized with respect to the posterior parameters $\theta_z$ and the likelihood parameters $\theta_x$.

In a vanilla variational autoencoder (VAE), the prior distribution over the latent variables 
is chosen to be a standard normal distribution:
\[
p(z) = \mathcal{N}(0, I).
\]
And, the approximate posterior distribution is assumed to be Gaussian with a diagonal covariance 
matrix, whose mean and standard deviation are parameterized by a neural network:
\[
q(z \mid x; \theta_z) = \mathcal{N}\bigl(\mu_{\theta_z}(x), \mathrm{diag}(\sigma_{\theta_z}^2(x))\bigr).
\]

In this work, we use a structured VAE framework where $z$ consists of a standard variational part, $z_v$ and a supervised part $z_s$. The prior distribution of the latent variables are,
\begin{eqnarray}
    p(z_v) &=& \mathcal{N}(0, I),\\
    p(z_s) &=& \mathcal{U}(-\infty,\infty).
\end{eqnarray}
And, the posterior distributions are,
\begin{eqnarray}
    q(z_v \mid x;\theta_{z_v}) &=& \mathcal{N}\bigl(\mu_{\theta_{z_v}}(x), \mathrm{diag}(\sigma_{\theta_{z_v}}^2(x))\bigr) \\
    q(z_s \mid x;\theta_{z_s}) &=& \delta\bigl(z_s-\mu_{\theta_{z_s}}(x)\bigr)
\end{eqnarray}
where $\mu_{\theta_{z_v}}(x)$, $\sigma_{\theta_{z_v}}^2(x)$ and $\mu_{\theta_{z_s}}(x)$ are the output of the encoder network.

In addition, the structured VAE assumes Gaussian likelihood for the data:
\begin{align}
p(x,y \mid z;\theta_x)
&= p(x \mid z;\theta_x)\,p(y \mid z_v) \notag \\
&= \mathcal{N}\!\left(
x \mid \mu_{\theta_x}(z),\, \sigma_x^2 I
\right)
\mathcal{N}\!\left(
y \mid z_s,\, \sigma_y^2 I
\right).
\end{align}

where the first term on the right hand side is similar to that in a vanilla VAE
where $\mu_{\theta_x}(z)$ is the output of the decoder network and $\sigma^2_x$ is typically assumed to be fixed; the second term corresponds to the supervised latent parameter constrained by data $y$. Under this assumption, the log-likelihood takes the form:
\begin{align}
\log p(x,y \mid z;\theta_x)
&=
-\frac{1}{2\sigma_x^2}
\lVert x - \mu_{\theta_x}(z) \rVert^2
-\frac{1}{2\sigma_y^2}
\lVert y - z_s \rVert^2
\notag \\
&\quad
-\frac{d_x}{2}\log\!\left(2\pi\sigma_x^2\right)
-\frac{d_y}{2}\log\!\left(2\pi\sigma_y^2\right).
\end{align}

where $d_x$ and $d_y$ are the dimensionality of $x$ and $y$.

Substituting this expression into the $\mathcal{L}_{ELBO}$ in Eq. (\ref{eq:elbo}), shows that maximizing the expected log-likelihood 
term is equivalent (up to an additive constant) to minimizing the loss function:
\begin{align}
-\mathcal{L}_{\mathrm{ELBO}}
&=
\frac{
\mathbb{E}_{q(z_v \mid x)}
\!\left[
\lVert x - \mu_{\theta_x}(z) \rVert^2
\right]
}{2\sigma_x^2}
+ \frac{
\lVert y - \mu_{\theta_{z_s}}(x) \rVert^2
}{2\sigma_y^2}
\notag \\
&\quad
+ \frac{1}{2}
\Bigl\|
\mu_{\theta_{z_v}}^{2}
+ \sigma_{\theta_{z_v}}^{2}
- \log \sigma_{\theta_{z_v}}^{2}
- 1
\Bigr\|.
\end{align}

with the expected mean squared error achieved by batch sampling of $z_v$ using the reparameterization trick:
\[
z_v = \mu_{\theta_{z_v}}(x) + \sigma_{\theta_{z_v}}(x) \odot \varepsilon,
\qquad
\varepsilon \sim \mathcal{N}(0, I),
\]
which enables differentiable, low-variance gradient-based optimization. The last term corresponds to the KL divergence 
$\mathrm{KL}\!\left(q(z_v \mid x_n;\theta_{z_v})\,\|\,p(z_v)\right)$, which involves two normal distributions and can be written analytically. The other KL divergence term involves supervised latent parameters 
$\mathrm{KL}\!\left(q(z_s \mid x_n;\theta_{z_s})\,\|\,p(z_s)\right)$ is a constant and can be neglected.

The VAE is an artificial neural network architecture that consists of the components shown in Fig. \ref{fig:VAE} \cite{VAE1} \cite{VAE2} \cite{VAEns3}. The neural network architecture and activation function at each layer used in this work is given in Tab. \ref{tab:vae_arch}. The goal of the encoder is to learn the probability distribution of the lower-dimensional latent variables \textit{z} given the input dataset \textit{X} discussed in Sec. \ref{subsec:vae-input}. This dimensionality reduction enables the network to compress complex, high-dimensional EOS information into a smooth latent manifold that captures the underlying physical correlations and variability in data. In doing so, it regularizes the learning process, mitigates overfitting to numerical artifacts, and facilitates an interpretable mapping between microphysical features of the EOS and macroscopic NS observables such as $M_{max}$ and $R_{1.4}$. A standard autoencoder directly outputs latent vectors from the encoder, while the VAE has a sampling layer composed of the mean $\mu$ and the log of the variance $\log(\sigma^2)$ calculated from the encoder. From this sampling layer, we draw a random sample $\varepsilon \sim \mathcal{N}(0, I)$ from the standard normal distribution and compute latent vectors using the transformation
$z = \mu + \exp\left(0.5 \cdot \log\sigma^{2}\right) \cdot \varepsilon$.
This formulation enables stochastic sampling from the latent Gaussian distribution while preserving differentiability with respect to $\mu$ and $\sigma$. Consequently, gradients can propagate through the sampling process during backpropagation, allowing the encoder to learn an appropriate latent distribution.
    
Additionally, the encoder learns a distribution $z_s$ of the supervised latent observables ($M_{max}$ and $R_{1.4}$) as a part of the latent layer for later use. The sampled latent vectors \textit{z} are then passed into the decoder, which increases dimensionality and computes a new probability distribution \textit{$\hat{X}$} of $z$. The output of the decoder is the reconstructed $c_s^2$ data with boundary conditions, which is used to compute the generated EOS, and is then passed into an algorithm that computes mass-radius (MR) curves for a given EOS using the TOV equations from Sec. \ref{subsec:TOV}. These MR values are stored in $\hat{Y}$. 

Throughout training, a total loss function is constructed to jointly optimize the reconstruction accuracy of the EOS, the supervised agreement between the predicted and true supervised latent observable values, and the Kullback-Leibler divergence that regularizes the latent space toward a standard normal prior, thereby balancing physical fidelity, predictive performance, and latent-space smoothness. The loss function used is given by 
\begin{equation}
\mathcal{L}
=
\mathcal{L}_{\mathrm{rec}}
+ \kappa\,\mathcal{L}_{\mathrm{sup}}
+ \eta\,\mathcal{L}_{\mathrm{KL}}.
\label{loss}
\end{equation}

\begin{align}
\mathcal{L}
&=
\overline{\lVert X - \hat{X} \rVert^2}
+ \kappa\,\overline{\lVert Y - y \rVert^2}
\notag \\
&\quad
+ \frac{\eta}{N\,d}
\sum_{n=1}^{N}
\sum_{i=1}^{d}
\left[
-\frac{1}{2}
\left(
1 + \log \sigma_{n,i}^{2}
- \mu_{n,i}^{2}
- \sigma_{n,i}^{2}
\right)
\right]
\label{loss_actual}
\end{align}

where $\mathcal{L}_{rec}$ is the reconstruction loss between the input and reconstructed data, $\mathcal{L}_{sup}$ is the reconstruction loss between the input supervised latent observables and those predicted by the encoder, $\mathcal{L}_{KL}$ is the Kullback-Leibler (KL) divergence, and $\eta, \kappa$ are weighting terms. The KL divergence is a measure of how different our prior $\sim \mathcal{N}(0, I)$ is from our probability distribution predicted by the encoder. We compute it for each latent dimension $i$ up to the total number of latent dimensions $d$, sum the contribution of each dimension, and take the mean by dividing by the number of samples in the batch $N$ multiplied by the number of latent dimensions. The KL divergence acts as a regularizing term, meaning that it can help coax the probability  distribution of the variational latent variables towards a Gaussian profile by penalizing it for deviations \cite{2023ApJ...950...77H}. Thus, a higher KL weighting $\eta$ forces the latent space distribution closer to a normal distribution. $\eta$ must be selected to be sufficiently small in order to minimize MAPE while avoiding posterior collapse, in which the latent variables become weakly informative and the decoder ignores the latent structure. Additionally, as $\eta$ goes to zero, we recover an autoencoder framework as we lose the probabilistic sampling nature of the VAE \cite{VAE1}.

\begin{table}
\caption{Dataset splits used for training and evaluation.}
\label{tab:data_splits}
\centering
\small
\setlength{\tabcolsep}{4pt}
\renewcommand{\arraystretch}{1.15}
\begin{tabular}{l c p{2.3cm} p{2.4cm}}
\hline
Split & Size & Purpose & During training \\
\hline
Training
& $\sim$70\%
& Fit model parameters
& Yes (updates weights) \\
Validation
& $\sim$15\%
& Hyperparameter and epoch selection
& No (evaluation only) \\
Test
& $\sim$15\%
& Final unbiased performance
& No (never used during training) \\
\hline
\end{tabular}
\end{table}

As outlined in Tab. \ref{tab:vae_arch}, the ReLU activation function is used in our work. The ReLU (Rectified Linear Unit) activation function allows positive values to pass through unmodified, while clipping negative values to zero \cite{agarap2019deeplearningusingrectified}. The output activation function could be better chosen to explicitly enforce causality, for example by using a sigmoid function. An activation function for the output layer did not affect our results of this work, but could be an interesting area of further testing and exploration.

\begin{table}
\caption{Neural network architecture.}
\label{tab:vae_arch}
\centering
\begin{tabular}{llcc}
\hline
Layer & Type & Neurons & Activation \\
\hline
Input        & ---    & 107 & --- \\
Layer 1        & Dense    & 128 & ReLU \\
Layer 2      & Dense  & 64  & ReLU \\
Layer 3      & Dense  & 64  & ReLU \\
Latent layer & Lambda & 4   & --- \\
Layer 4      & Dense  & 64  & ReLU \\
Layer 5       & Dense  & 128 & ReLU\\
Output (a)      & Dense  & 101 & softnegative \\
Output (b)      & Dense  & 6 & ---\\
Output & Concatenate  & 107 & ---\\
\hline
\end{tabular}
\end{table}

The primary machine learning system we use is the Python package TensorFlow \cite{2016arXiv160508695A}. We use a batch size of 64 and a learning rate of $\alpha=.0001$. Our input data consisting of the array discussed in Sec. \ref{subsec:vae-input}, as well as our supervised latent observables $M_{max}$ and $R_{1.4}$, undergo a transformation using the StandardScaler given by Scikit-learn. This standardizes the data by removing the mean and scaling to unit variance.  
For the selected dimension of the latent space, we refer to the mean absolute percentage error (MAPE) heatmap shown in Fig. \ref{fig:decoder_mape}. The MAPE is computed by the equations shown in Fig. \ref{fig:VAE}. The heatmap shows the MAPE computed between the decoder-predicted values of our supervised latent observables and the encoder-learned supervised latent observables that are part of the test set. This MAPE is particularly important since it measures how faithfully the decoder preserves the mapping between the latent variables and physically observable NS properties. A low decoder MAPE therefore indicates that the latent space encodes sufficient information to accurately reproduce the target observables, although it does not by itself imply physical interpretability or disentanglement of individual latent dimensions. We train a model and compute the MAPE for each shown combination of hyperparameter $\eta$ and $\kappa$, and for latent dimensions 1-4. Each tile in the heatmap represents a different training run using the corresponding hyperparameters and dimensionality. We use this heatmap to select the model that we will analyze further in Sec. \ref{sec: results}. We choose a combination with a low MAPE that is similar for both supervised latent observables, ensuring that the latent space encodes both quantities with similar fidelity. For the MAPE corresponding to the $M_{max}$, a value of $.10\%$, and $.15\%$ for the value corresponding to $R_{1.4}$. This corresponds to hyperparameter values of $\eta=0.001$ and $\kappa=2$, with a latent dimensionality of one. The selected combination is outlined with a red box in Fig. \ref{fig:decoder_mape}. The latent dimension choice can be made by inspecting the latent space, as we do in Sec. \ref{subsec: latent}.

\section{Results}\label{sec: results}
\begin{figure*}[t]
    \centering

    \begin{subfigure}{\textwidth}
        \centering
        \includegraphics[width=\textwidth]{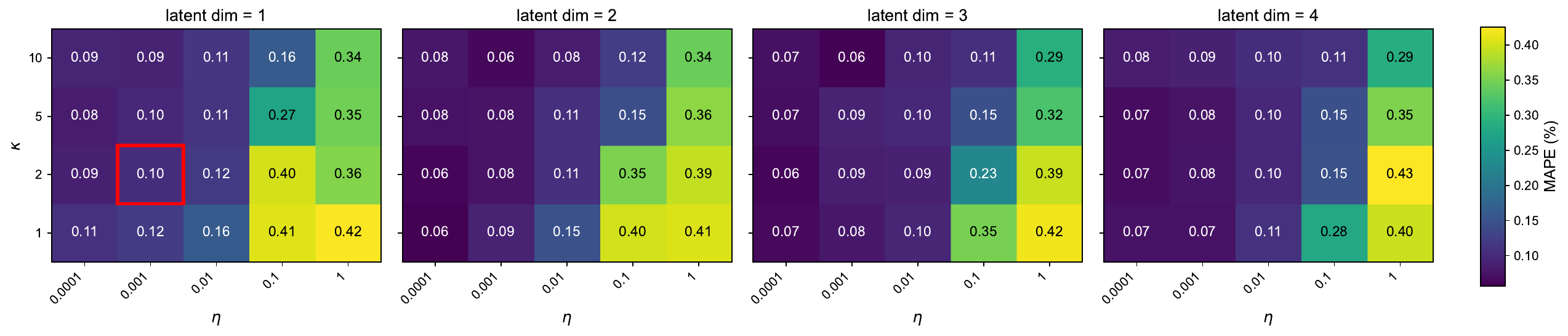}
        \caption{MAPE heatmap for the maximum mass $M_{\max}$}
        \label{fig:mmax_mape}
    \end{subfigure}

    \par\vspace{0.75\baselineskip}

    \begin{subfigure}{\textwidth}
        \centering
        \includegraphics[width=\textwidth]{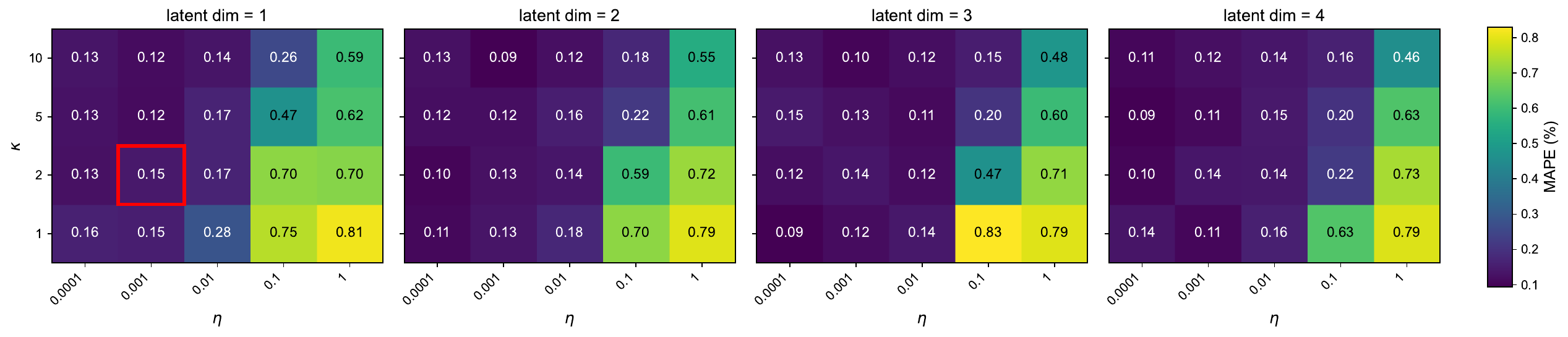}
        \caption{MAPE heatmap for the radius of a $1.4\,M_{\odot}$ NS ($R_{1.4}$)}
        \label{fig:r14_mape}
    \end{subfigure}

    \caption{Decoder MAPE comparison for both supervised latent observables defined in Fig. \ref{fig:VAE}. The MAPE is computed for each combination of latent dimensionality, $\kappa$, and $\eta$ used during training. The hyperparameter and dimensionality combo selected for further analysis is outlined with a red box.}
    \label{fig:decoder_mape}
\end{figure*}

After training is completed, we now have a fully generative VAE model that we can sample from to generate EOS. The EOS can be fully controlled using a single latent parameter $z_0$. The decoder, conditioned by the supervised latent observables $M_{max}$ and $R_{1.4}$, is able to make sense of the single latent parameter and convert $z_0$ into meaningful EOS. Thus, we can now mimic a ten-parameter Skyrme model using only two supervised latent observables and one latent variable, for a total of three parameters. In this section, we will analyze the model indicated in Fig. \ref{fig:decoder_mape} in detail.

\begin{figure}[ht!]
    \centering
    \includegraphics[width=\columnwidth]{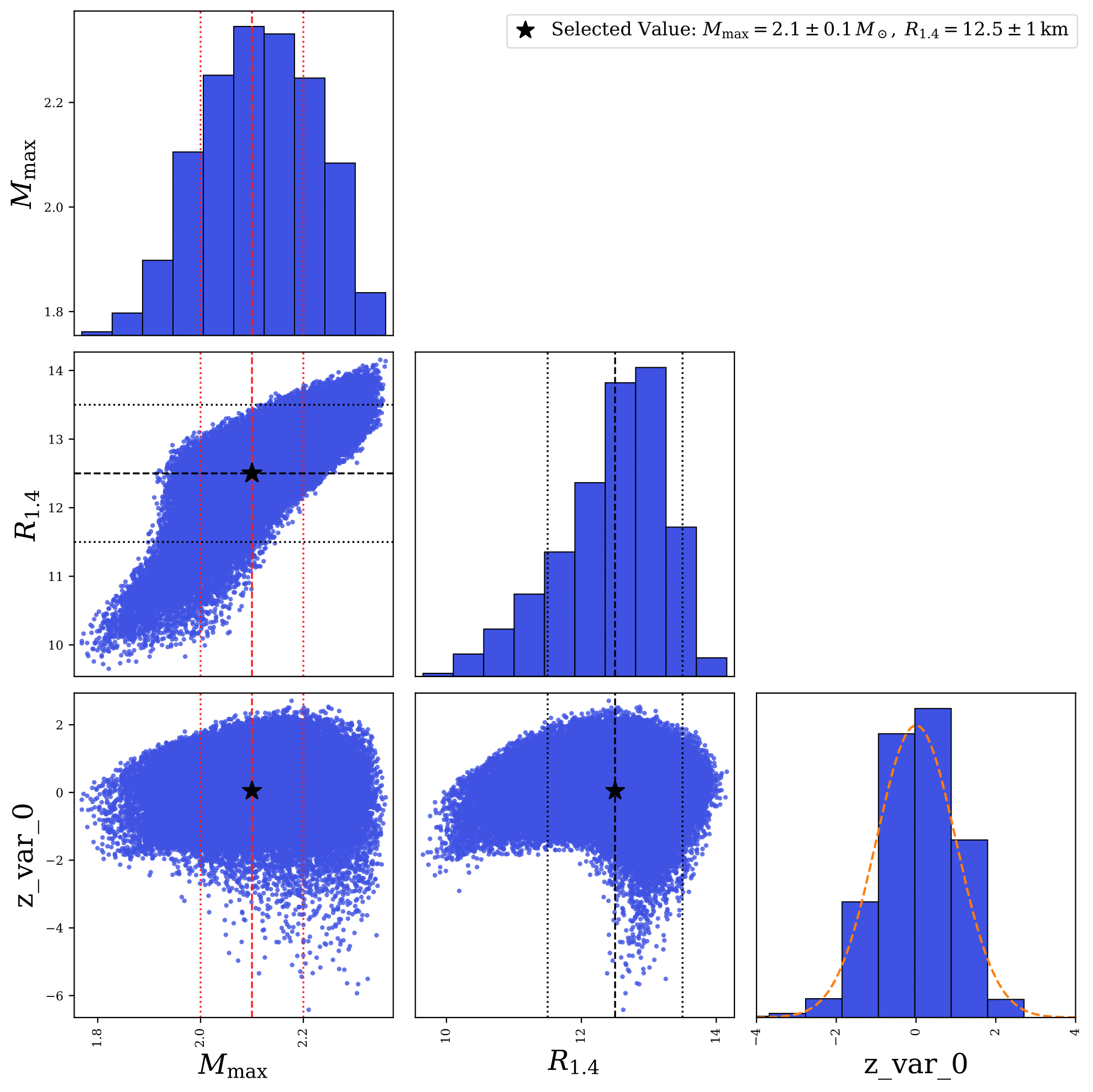}
    \caption{Pairwise distributions for the supervised and latent variables for the test dataset. The selected values of $M_{\max}$ and $R_{1.4}$, along with the intervals used to probe latent space sensitivity in the reconstructed EOS, are indicated by the dashed lines. For the latent variable $z_0$, a standard normal distribution (orange) is overlaid on the histogram.}
    \label{fig:latentspace}
\end{figure}

\subsection{Latent Space}\label{subsec: latent}

Fig. \ref{fig:latentspace} shows the latent space distribution of the test dataset, corresponding to the parameter space that the decoder can sample from. This trained VAE model has one latent dimension and two supervised latent observables, $M_{\max}$ and $R_{1.4}$, with hyperparameters $\eta = 0.001$ and $\kappa = 2$. This configuration was selected based on the criteria discussed in Sec. \ref{subsec:SVAE}. We choose the number of latent variables $z_i$ based on the minimum dimensionality required to capture nontrivial, physically meaningful variability in the EOS. Latent dimensions that collapse to near-linear manifolds or exhibit strongly non-Gaussian, degenerate distributions are indicative of redundant or inactive modes and are therefore excluded.
The supervised latent observables exhibit a strong linear correlation as expected, quantified by the Pearson coefficient $r(M_{\max},\, R_{1.4}) = 0.8117$, computed over the full test set. The latent variable $z_0$ in Fig. \ref{fig:latentspace} follows an approximately Gaussian distribution as imposed by the KL divergence. When projected against the supervised latent observables, we see a smooth, nonlinear correlation, indicating that this latent dimension has learned a physically meaningful mode of variation in the EOS.

\subsection{Decoded EOS and Mass-Radius Curves}
To test the generative capabilities of the trained VAE model, we vary individual latent parameters and supervised latent observables over their posterior-supported ranges while fixing all other dimensions at their mean values. Each perturbation is decoded into a full EOS, which is then propagated through the TOV equations to produce MR relations. This procedure allows us to verify that smooth, physically interpretable variations in latent space correspond to continuous, stable, and astrophysically consistent deformations of NS structure, thereby validating both the physical expressivity and generative robustness of the learned model.

\begin{figure*}[ht]
    \centering
    \includegraphics[width=\textwidth, clip, trim=0 0 0 0]{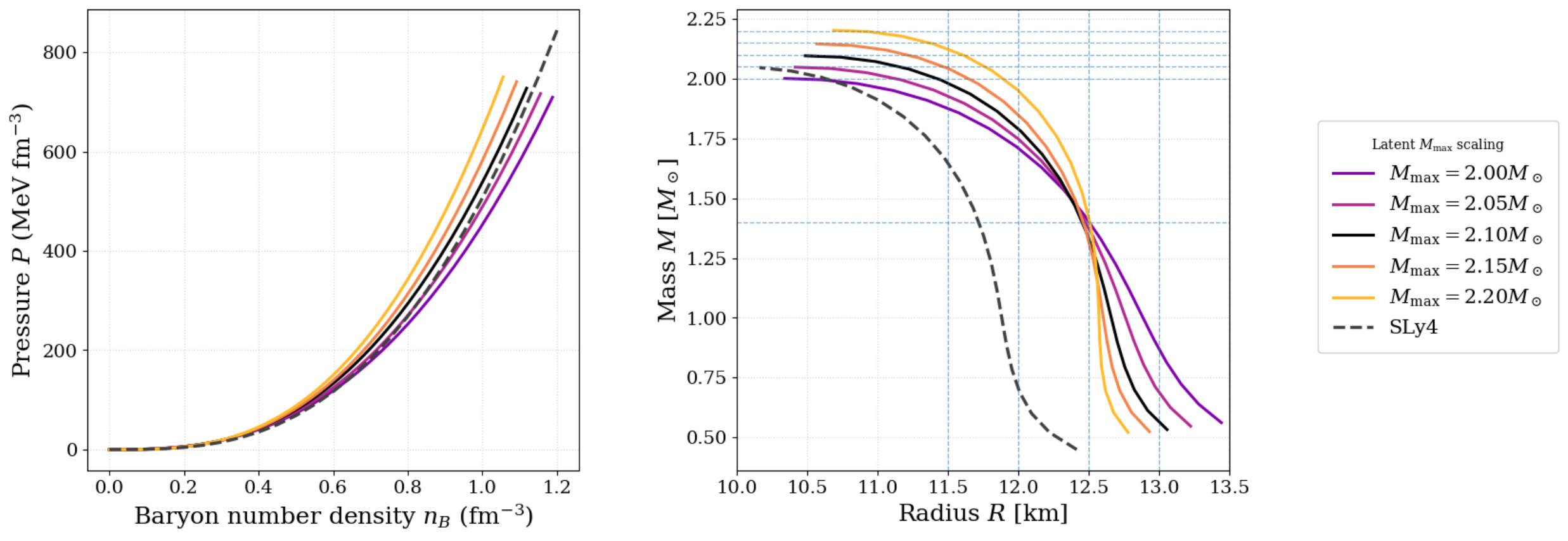}
    \caption{EOS $P(n_B)$ (left panel) generated by varying the supervised latent parameter 
    $M_{\max}$ about a central value of $2.1\,M_\odot$, while holding 
    $R_{1.4}$ and $z_0$ fixed. For each chosen $M_{\max}$, the corresponding EOS is 
    decoded from the VAE latent space and evaluated over a common pressure grid. 
    For comparison, we also show the SLy4 EOS (dashed). The right panel shows the 
    mass--radius curves computed from each corresponding EOS.}
    \label{fig:EOS_sensitivity_Mmax_P_vs_nB}
\end{figure*}

Fig. \ref{fig:EOS_sensitivity_Mmax_P_vs_nB} shows the first such perturbation. We fix the latent variable $z_0$ and the supervised latent observable $R_{1.4}$ while varying the other supervised latent observable, $M_{max}$ about the central selected value as indicated in Fig. \ref{fig:latentspace}. For the EOS $P(n_B)$, the decoded equations of state remain tightly clustered at low baryon densities, indicating that the crust and outer-core behavior is largely unaffected by changes in the supported maximum mass. At higher densities, however, the EOS curves begin to separate noticeably, reflecting increasing sensitivity to the assumed high-density physics. We find that the decoded equations of state become systematically stiffer at high densities as $M_{max}$ is increased, consistent with the physical expectation that greater pressure support at supranuclear densities is required to stabilize more massive NSs against gravitational collapse \cite{2001ApJ...550..426L}. The exact SLy4 EOS and corresponding MR curve are plotted for reference.

\begin{figure*}[ht]
    \centering
    \includegraphics[width=\textwidth]{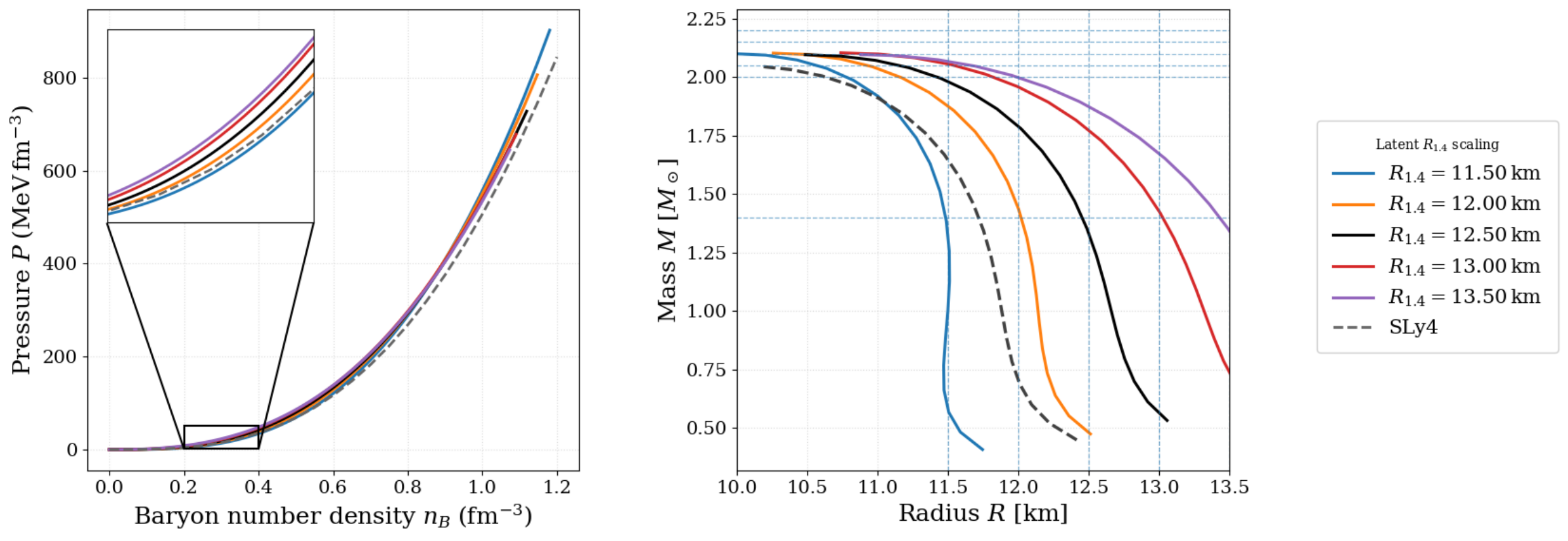}
    \caption{Same as Fig. ~\ref{fig:EOS_sensitivity_Mmax_P_vs_nB}, except the supervised latent parameter $M_{\max}$ and the latent variable $z_0$ are held fixed while varying $R_{1.4}$ about a central value of $12.5\,\mathrm{km}$.}
    \label{fig:EOS_sensitivity_R14_P_vs_nB_zoom}
\end{figure*}

\begin{figure*}[ht!]
    \centering
    \includegraphics[width=\textwidth]{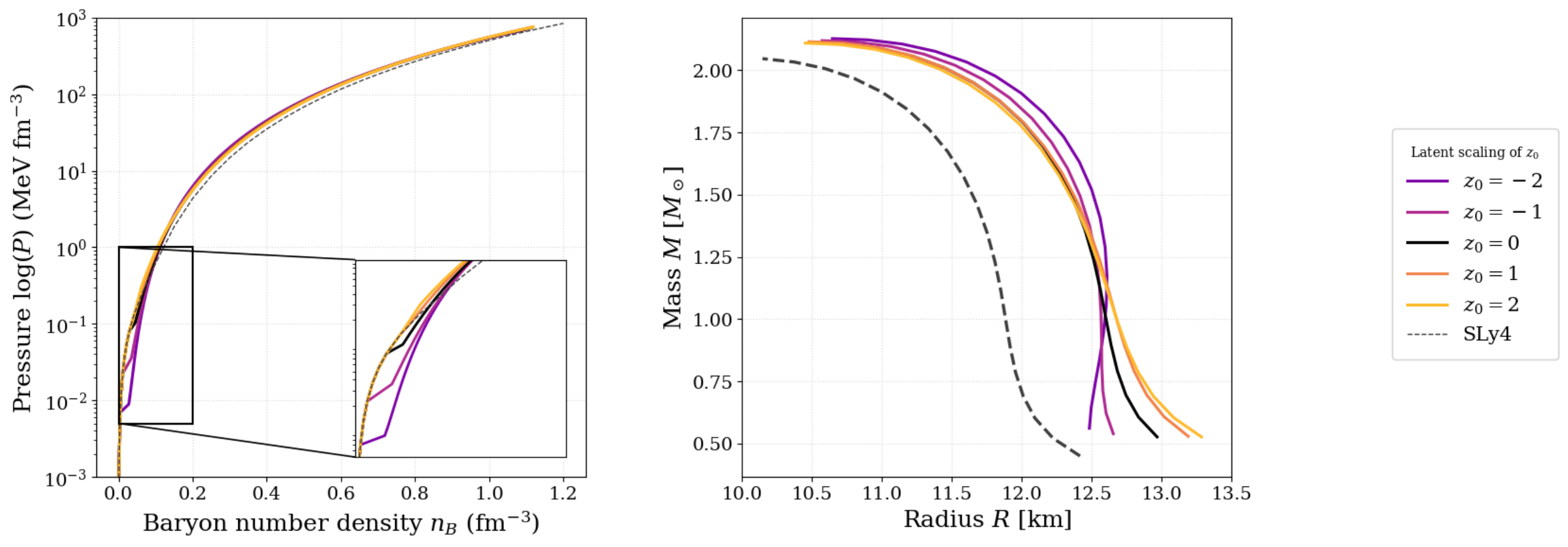}
    \caption{Same as Fig. ~\ref{fig:EOS_sensitivity_Mmax_P_vs_nB} and ~\ref{fig:EOS_sensitivity_R14_P_vs_nB_zoom}, except the supervised latent parameter $M_{max}$ and $R_{1.4}$ are held fixed while $z_0$ is varied about a central value of 0.} 
    
    \label{fig:EOS_sensitivity_latent}
\end{figure*}

In Fig. \ref{fig:EOS_sensitivity_R14_P_vs_nB_zoom}, we perform the same procedure as in Fig. \ref{fig:EOS_sensitivity_Mmax_P_vs_nB}, but we instead fix $M_{max}$ and $z_0$ while varying $R_{1.4}$ (the supervised latent observable corresponding to the radius of a $1.4 \: M_{\odot}$ NS). The EOS remain coherent at higher densities corresponding to the core of a NS, but noticeably diverge at the low-density region highlighted in the upper-left panel. This behavior is consistent with the well-known sensitivity of NS radii to the pressure near and just above nuclear saturation density, where variations in the EOS primarily influence the stellar envelope rather than the core \cite{2013ApJ...773...11H}.

We do the same procedure as above in Fig. \ref{fig:EOS_sensitivity_latent}, but instead varying the latent variable $z_0$ while fixing $M_{max}$ and $R_{1.4}$ and decoding the corresponding EOS. This variation produces EOS that are very similar, with variations at baryon density $n_B \lesssim 0.1$ fm$^{-3}$. Differences in the low-density EOS directly translate into variations in NS radii, particularly for lower-mass stars, indicating that $z_0$ primarily encodes information relevant to the intermediate-to-low density structure of NS.

An interesting feature to note is the EOS crossing point that occurs at low densities around nuclear saturation density in Fig. \ref{fig:EOS_sensitivity_Mmax_P_vs_nB} (more apparent in log-scaled space), and the corresponding MR curve crossing that occurs around a mass of $1.4$ $M_{\odot}$ and a radius of $12.5$ km. These features are likely correlated and reflect the sensitivity of $R_{1.4}$ to the EOS at near saturation density discussed above. This crossing is clear to see in the EOS curves as shown in Fig. ~\ref{fig:EOS_sensitivity_R14_P_vs_nB_zoom}.

\subsection{Decoded vs. SLy4 EOS Comparison}
\begin{figure*}[t]
    \centering
    \includegraphics[width=\textwidth, clip, trim=0 0 0 0]{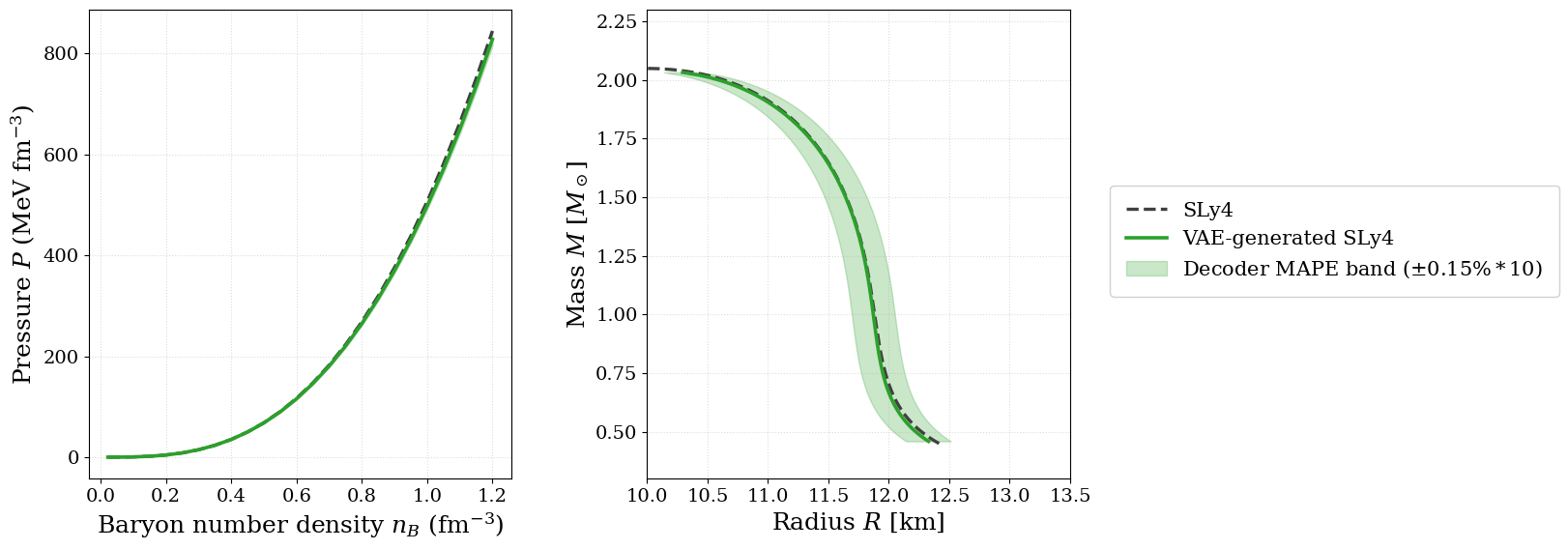}
    \caption {Equation of state $P(n_B)$ and MR curves generated by fixing the values of 
    $M_{\max}$ and $R_{1.4}$ to the SLy4 values of $2.046$ $M_\odot$ and $11.717$ km, respectively and decoding the EOS. Shaded regions denote the fractional tolerance band obtained by scaling the decoded EOS by the respective decoder MAPEs. The exact SLy4 model EOS and MR curve is shown for reference.}
    \label{fig:sly4comp}
\end{figure*}

We now aim to show that the VAE framework is capable of accurately reproducing the SLy4 EOS by directly selecting and passing to the decoder the supervised latent observables $M_{max}=2.046$ $M_\odot$ and $R_{1.4}=11.717$ km, values calculated with the SLy4 parameterization. Fig. \ref{fig:sly4comp} shows the EOS and corresponding MR curves for SLy4, and the VAE-generated EOS made by explicitly passing in the above values of $M_\odot$ and $R_{1.4}$ as conditioning points to the decoder. For the value of the latent parameter $z_0$, we use the scaled latent mean value of $z_0=.0487$. We also use the computed value of the decoder MAPE for $R_{1.4}$ of $0.15\%$ (as given in Fig. \ref{fig:decoder_mape}) to construct a symmetric fractional tolerance band about the VAE-generated MR curve. This acts as a visual indicator of the high accuracy the VAE achieves when attempting to decode the SLy4 EOS. In particular, the VAE decodes a value of $M_{max}=2.043\pm0.002$ $M_{\odot}$ and $R_{1.4}=11.715\pm.0176$ km, in very close agreement to the exact SLy4-calculated values of $2.046$ $M_{\odot}$ and $11.717$ km. We scale up the tolerance band in Fig. ~\ref{fig:sly4comp} by a factor of 10 for visibility.

\section{Conclusion}
In this work, we create a VAE framework to generate new candidate NS EOS. We show that our model is capable of taking theoretical 10-parameter Skyrme EOS as input EOS data, compressing it down to a single latent dimension $z_0$ and two supervised latent observables $M_{max}$ and $R_{1.4}$. We can then use this reduced dimension latent space to control the generation of physically consistent NS EOS. By varying the latent space values, we demonstrated that our decoded EOS remain smooth, causal, and thermodynamically stable across the full density range of interest, while exhibiting controlled and physically interpretable variations in both macroscopic NS observables and high-density EOS behavior. In particular, systematic changes in the supervised and variational latent parameters produce coherent shifts in the stiffness of the EOS at different density ranges and the corresponding mass–radius relations at different mass ranges. Together, these results demonstrate that the latent representation learned by the VAE provides a compact and physically meaningful parameterization of the NS EOS, enabling efficient exploration of EOS sensitivity within observationally and theoretically admissible bounds.

\section{Future Work}
Future work should include a comparison of the speed increases of this VAE approach as compared to traditional methods of creating new EOS, to cement its robustness as an alternative, if not improved, method for NS EOS studies. One of the primary advantages of using a decoder network to perform Markov chain Monte Carlo sampling of latent space parameters for Bayesian analysis involving NS observations is the decoder can generate large numbers of candidate EOS realizations essentially “on demand”. This potentially enables faster exploration of EOS uncertainty while maintaining physically motivated constraints embedded in the training data and EOS construction procedure, which needs to be explored in future works.

This framework supports the integration of other theoretical nuclear models into the input data used to train the model. Naturally, adding EOS and boundary conditions derived from models other than Skyrme, such as the Relativistic Mean-Field model, would potentially allow a more thorough exploration of the theoretical EOS parameter space. In addition, scenarios involving first-order phase transitions, in contrast to the relatively smooth EOS functions considered here, should be explored. Such transitions under Maxwell construction between various quark models such as the MIT bag model and hadronic model EOS, naturally allow the sound speed to approach zero and introduce discontinuities in the EOS \cite{Lopes_2021} \cite{Blacker_2023}. These more sophisticated possibilities may be particularly well-suited for application of the VAE framework proposed in this work.

\section*{Acknowledgements}
This research is supported by N3AS's National Science Foundation award No. 2020275, and a scholarship from the Mary Gates Endowment for Students. We thank the Institute for Nuclear Theory at the University of Washington for its support.

\printbibliography

\end{document}